\documentclass[11pt,twoside]{article}

\usepackage{asp2006}
\usepackage{epsf}
\usepackage{lscape}

\begin{document}
\title{Post-AGB stars in the AKARI survey}   
\author{Natasza Si\'{o}dmiak$^1$, Nick Cox$^2$, Ryszard Szczerba$^1$, Pedro
Garc\'{i}a-Lario$^2$ }   
\affil{
$^1$ N. Copernicus Astronomical Center, Toru\'{n}, Poland \\
$^2$ Herschel Science Centre, European Space Astronomy Centre, Spain}    

\begin{abstract} 
Obscured by their circumstellar dusty envelopes post-AGB stars emit a large
fraction of their energy in the infrared and thus, infrared sky surveys like
IRAS were essential for discoveries of post-AGBs in the past. Now, with
the AKARI infrared sky survey we can extend our knowledge about the late
stages of stellar evolution. The long-term goal of our work is to define new
photometric criteria to distinguish new post-AGB candidates from the AKARI
data.

We have cross-correlated the Toru\'{n} catalogue of Galactic post-AGB and
related objects with the AKARI/FIS All-Sky Survey Bright Source Catalogue
(for simplicity, hereafter AKARI). The scientific and technical aspects of
our work are presented here as well as our plans for the future. In
particular, we found that only 9 post-AGB sources were detected in
all four AKARI bands. The most famous objects like: Red Rectangle, Egg
Nebula, Minkowski's Footprint belong to this group. From the technical point
of view we discuss positional accuracy by comparing (mostly) 2MASS
coordinates of post-AGB objects with those given by AKARI; flux reliability
by comparing IRAS 60 and 100 $\mu$m fluxes with those from AKARI-N65 and
AKARI-90 bands, respectively; as well as completeness of the sample as a
function of the IRAS fluxes.
\end{abstract}


\section{Post-AGB sample}   
Post-Asymptotic Giant Branch (post-AGB) objects are rapidly evolving stars
of low and intermediate initial mass (0.8-8 M$_\odot$) in the transition phase
between AGB and planetary nebulae (PN). This phase is very short (of the
order of 1000 yrs) but yet very important and still poorly understood. The
departure from spherical symmetry or changes in surface chemical composition
during AGB are still studied using post-AGBs in order to understand the late
stages of stellar evolution.

Analyzed post-AGBs were taken from the very likely part (346 objects) of the
Toru\'{n} catalogue of Galactic post-AGB and related objects (\citet{sz09} - 
http://www.ncac.torun.pl/postagb2/). Sources were cross-correlated with the
AKARI catalogue. We found matches for 144 objects within 30$\arcsec$ radius
from the reference coordinates of post-AGB objects (usually 2MASS
coordinates), and all of them with good photometry at 90$\mu$m.

\section{Statistics}
\begin{figure}[!ht]
\begin{minipage}[b]{0.5\linewidth}
   \centering
   \resizebox{0.8\hsize}{!}{
     \includegraphics*{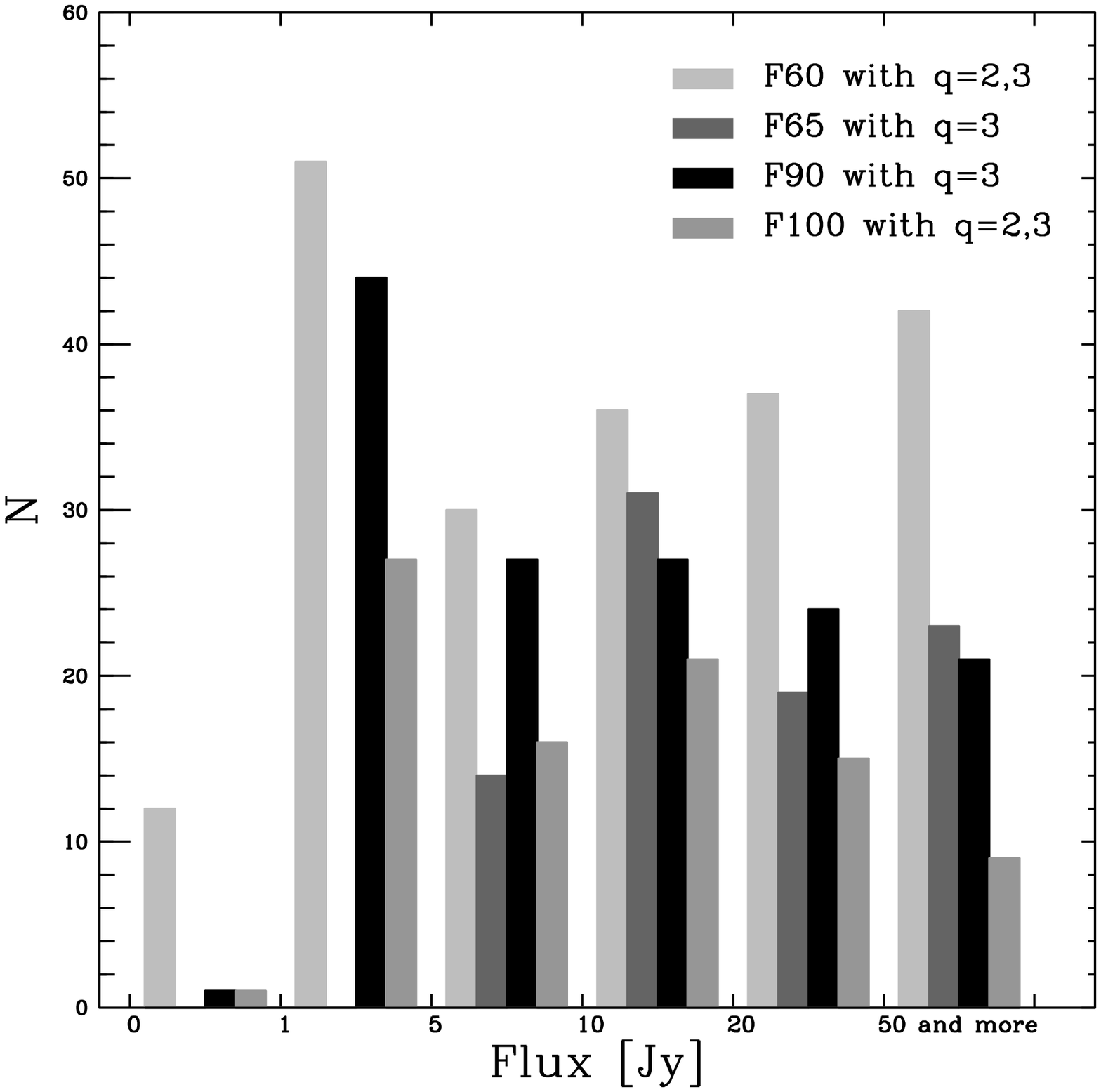}
  }
\caption{IRAS and AKARI sources in Toru\'{n} post-AGB catalogue.}
\label{compsam}
\end{minipage}
\hspace{0.1cm}
\begin{minipage}[b]{0.5\linewidth}
   \centering
   \resizebox{0.8\hsize}{!}{
     \includegraphics*{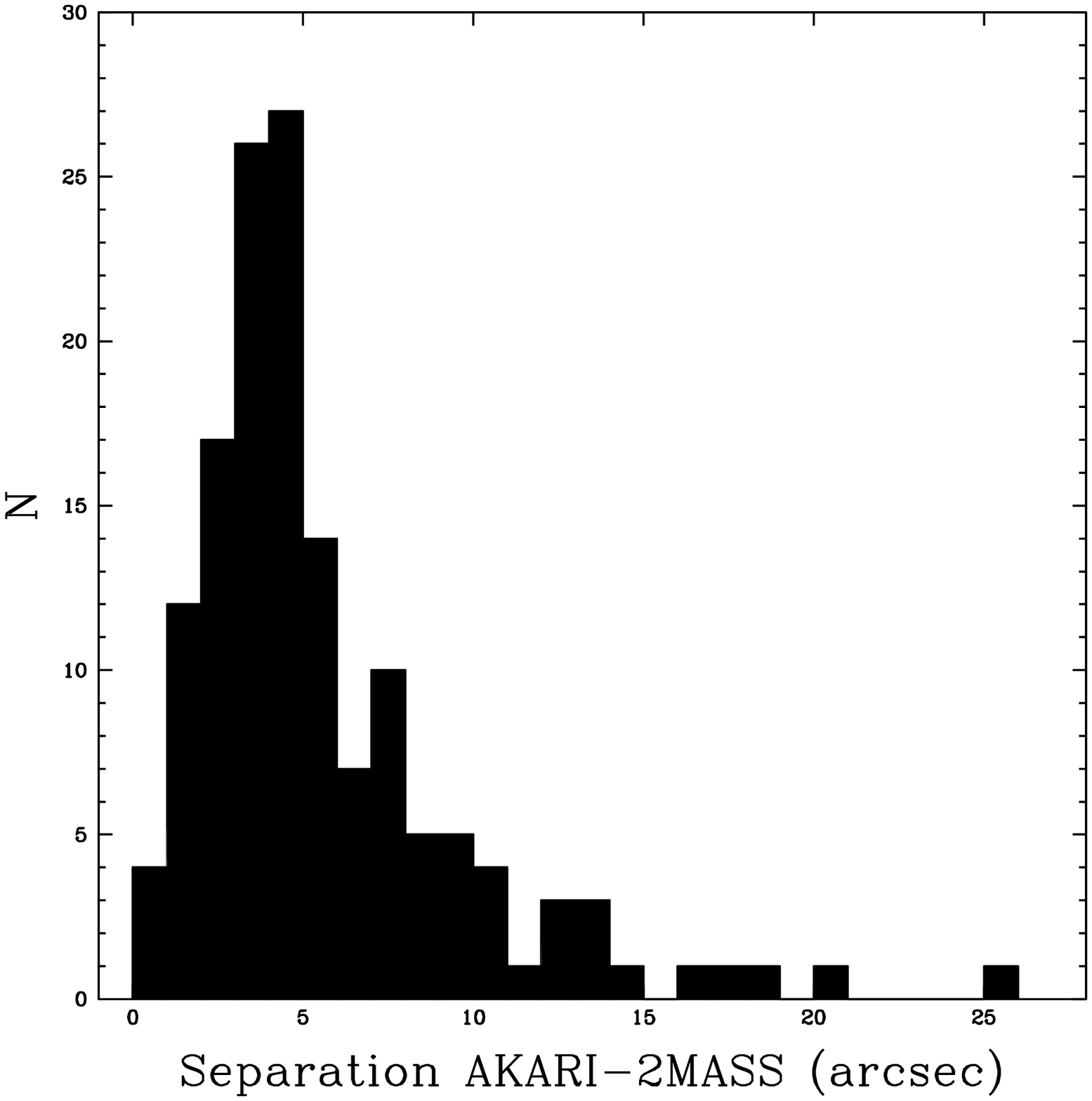}
  }
\caption{Separation of AKARI and 2MASS counterparts of post-AGB objects.}
\label{sep}
\end{minipage}
\end{figure}

There are 256 sources with IRAS counterparts in Toru\'{n} catalogue. Among 144
post-AGB objects found in AKARI, 143 have IRAS counterparts. This
implicates that ~56\% of post-AGBs with IRAS counterparts have AKARI
counterparts. There is a lack of AKARI sources with F65$<$5Jy when comparing
to IRAS sources. The completeness of our sample as a function of fluxes is
shown in Figure~\ref{compsam}.

The positional accuracy is in most cases in agreement with 2MASS
coordinates. For 139 out of 144 objects the AKARI-2MASS separation is below
15$\arcsec$ and for 127 of them below 10$\arcsec$ (Figure~\ref{sep}). Since
the IRAS coordinates differ from 2MASS ones, the AKARI-IRAS
separation is slightly different and resulting in 126/143 objects with the
separation below 15$\arcsec$ and 107/143 below 10$\arcsec$.

\begin{figure}[!ht]
\begin{center}
   \resizebox{0.9\hsize}{!}{
	\plottwo{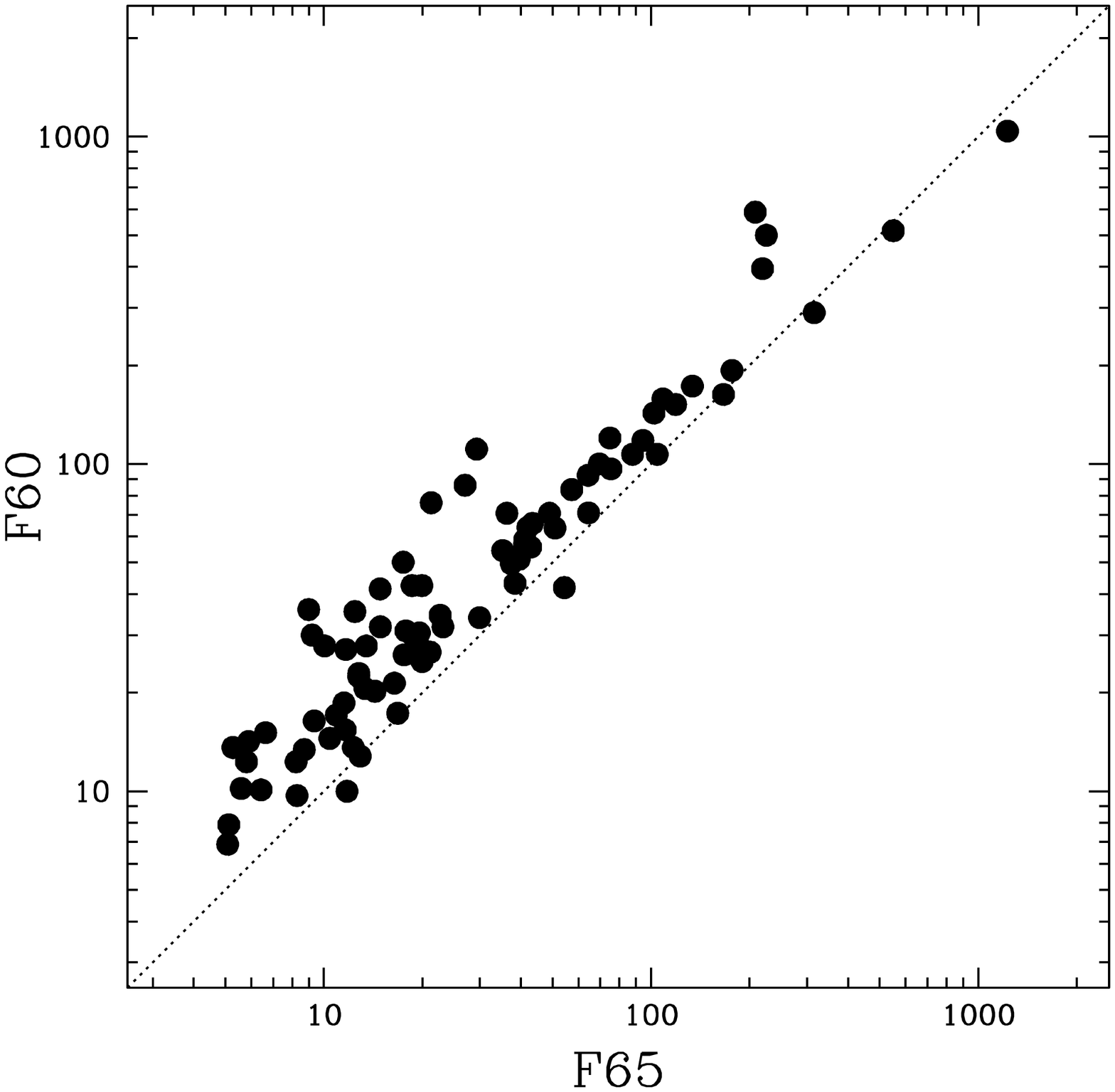}{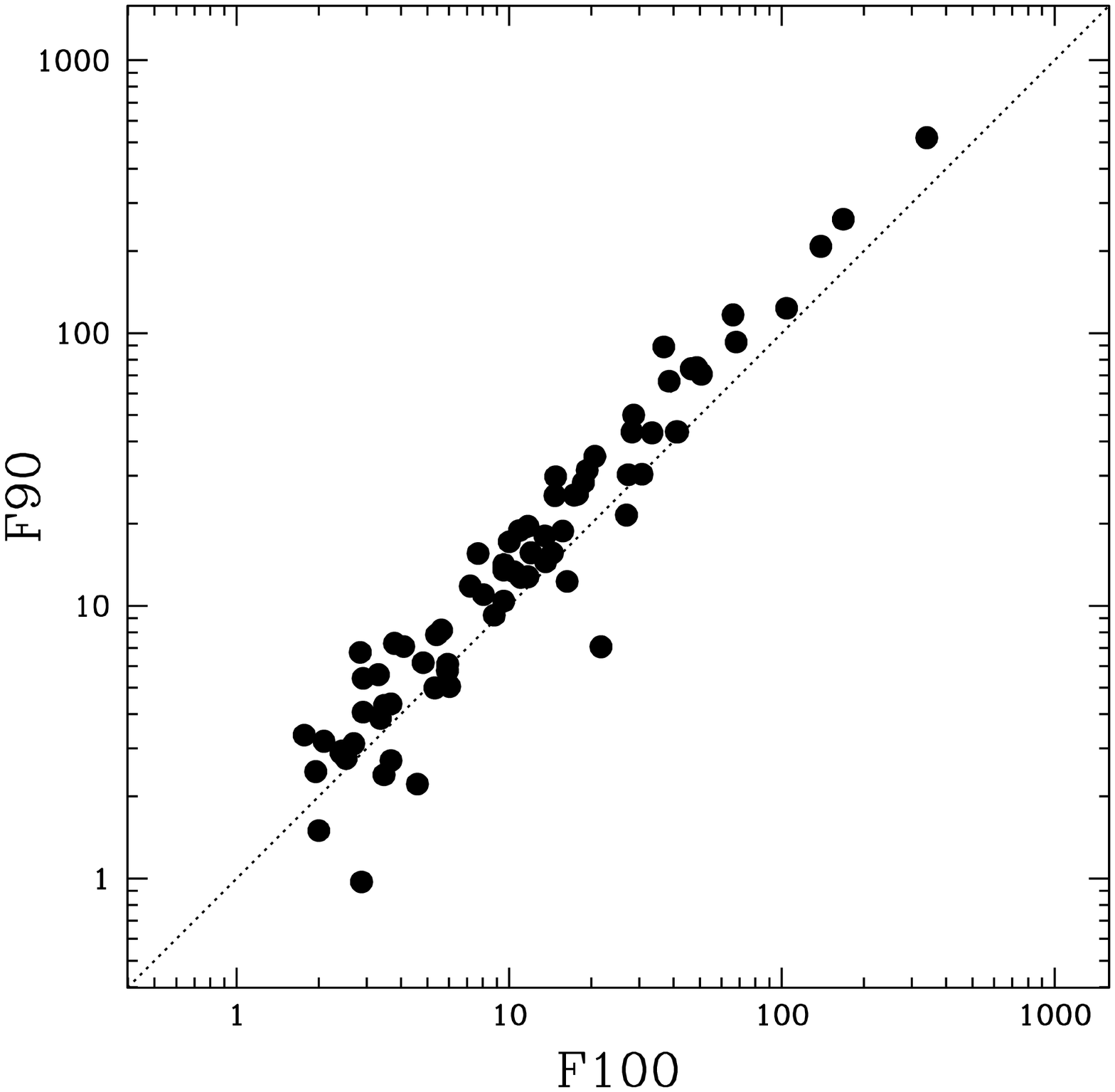} 
  }
\end{center}
\caption{The comparison of AKARI and IRAS fluxes (only with good quality). Dotted
line shows 1:1 relation.}
\label{akir}
\end{figure}

The comparison of AKARI and IRAS fluxes F60 vs. F65 and F90 vs. F100 is
shown in Figure~\ref{akir}. The correlations are roughly linear (restriction
to objects with cirrus emission less than 100 MJy/sr does not improve shown
relation). In general, post-AGB objects show maximum of their infrared
emission between 25 and 60$\mu$m and thus their flux is smaller toward
longer wavelengths, e.g. sources show less flux at 65$\mu$m than at 60$\mu$m
and, consequently, less flux at 100$\mu$m than at 90$\mu$m.

\begin{figure}[!ht]
\begin{center}
   \resizebox{0.47\hsize}{!}{
     \includegraphics*{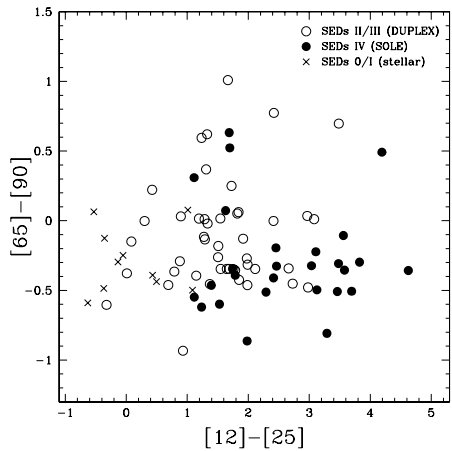}
  }
\end{center}
\caption{AKARI [65]-[90] vs IRAS [12]-[25] for post-AGBs. Different symbols
code types of SEDs (morphology class).}
\label{sole}
\end{figure}

According to \citet{ueta00} and \citet{sio08} we can divide
post-AGB sources into 3 groups depending on their morphology and spectral
energy distribution (SED): DUPLEX with SEDs of class II and III (classes by
\citet{van89}), SOLE with SEDs of class IV and stellar objects with SEDs of
class I or 0. Figure~\ref{sole} shows color-color diagram for post-AGB
objects with different SED classes. Stellar objects (crosses) have small
values of [12]-[25] characteristic for sources with no envelope. DUPLEX
objects (circles) have in general larger values of [65]-[90] than SOLEs
(black dots) as a consequence of more (cold) dust in their envelopes.

\section{Detections at 140 \&  160$\mu$m}
\begin{figure}[!ht]
\begin{center}
   \resizebox{0.8\hsize}{!}{
     \includegraphics*{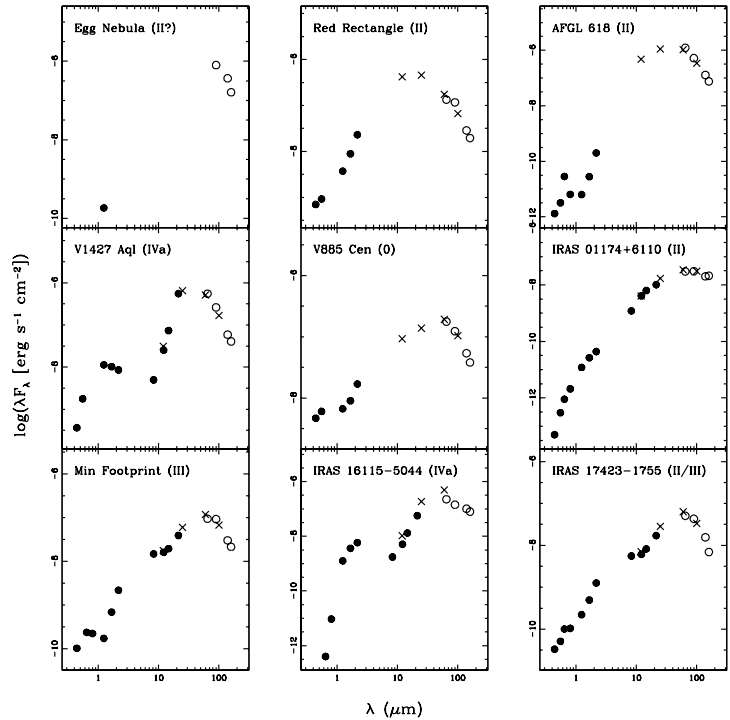}
  }
\end{center}
\caption{SEDs of post-AGBs detected in all AKARI bands (crosses –-- IRAS,
open points –-- AKARI photometry). SED classes are added.} 
\label{seds}
\end{figure}

At longer wavelengths there are 9 objects from our sample detected at both
140 and 160$\mu$m and additionally 25 sources detected only at 140$\mu$m.
SEDs of 9 post-AGBs are displayed in Figure~\ref{seds}. In general, objects
detected at 140 and/or 160$\mu$m are the brightest ones in our sample and
very dusty at the same time (e.g. Red Rectangle, AFGL 618, Egg Nebula).

\section{Future work}
[65]-[90] and/or [140]-[160] AKARI colors in combination with other infrared
measurements will be used to select new post-AGB candidates. Newly
discovered sources will be included in the Toru\'{n} catalogue of post-AGB
objects, together with all the available astrometric, photometric and
spectroscopic data.

Having AKARI fluxes we will be able to (better) constrain the dust
temperature of post-AGB objects.  We also hope to obtain AKARI images of
analyzed sources and thus get closer to understanding the late stages of
stellar evolution.

\acknowledgements 
Based on observations with AKARI, a JAXA project with the participation of
ESA.
This work has been supported by grant N203 0661 33 of MNiSW of Poland.
R.Sz. and P.G-L. acknowledge support from the Faculty of the European Space
Astronomy Centre (ESAC).


\end{document}